\documentclass[apj]{emulateapj}
\usepackage{bm}
\usepackage{amsmath}
\usepackage{amsfonts}
\usepackage{amsthm}
\usepackage{amssymb}
\usepackage{xspace}
\usepackage{graphicx}
\usepackage{graphics}
\usepackage{hyperref}

\newcommand{\be}{\begin{equation}}
\newcommand{\ee}{\end{equation}}
\newcommand{\eq}[1]{Equation~(\ref{eq_#1})}
\newcommand{\lcdm}{$\Lambda$CDM\xspace}
\newcommand{\kms}{{\rm km\,s^{-1}}}
\newcommand{\msun}{{\rm M}_{\odot}}
\newcommand{\lsun}{{\rm L}_{\odot}}
\newcommand{\msunyr}{{\rm M}_{\odot}\rm yr^{-1}}

\newcommand{\x}{\mathbf{x}}
\newcommand{\ha}{H{\sc\,i}\xspace}

\newcommand{\Mh}{M_{\rm h}}
\newcommand{\Rh}{R_{\rm h}}
\newcommand{\Vh}{V_{\rm h}}
\newcommand{\Jh}{J_{\rm h}}
\newcommand{\jh}{j_{\rm h}}
\newcommand{\Ms}{M_{s}}
\newcommand{\Mg}{M_{\rm g}}

\newcommand{\Mknee}{\Ms^\ast}
\newcommand{\Js}{J_s}
\newcommand{\js}{j_s}
\newcommand{\jg}{j_{\rm g}}

\newcommand{\Ss}{\Sigma_s}

\newcommand{\re}{R_{\rm e}}

\newcommand{\jobs}{j_{s,\rm obs}}
\newcommand{\jextra}{j_{s,\rm tot}}
\newcommand{\incl}{\eta}

\newcommand{\fg}{f_{\rm g}}
\newcommand{\Qm}{\overline{Q}}
\newcommand{\citex}[1]{ \citep{#1}}
\usepackage{color}
\definecolor{mycolor}{rgb}{0,0,1.0}
\newcommand{\link}[1]{\textcolor{mycolor}{#1}} 

\newcommand{\new}[1]{#1} 

\begin{document}

\title{Low angular momentum in clumpy, turbulent disk galaxies}



\author{\hspace{-1.0cm}Danail Obreschkow$^1$, Karl Glazebrook$^2$, Robert Bassett$^2$, David B. Fisher$^2$, Roberto G. Abraham$^3$, Emily\\\hspace{-0.8cm}Wisnioski$^4$, Andrew W. Green$^5$, Peter J. McGregor$^6$, Ivana Damjanov$^7$, Attila Popping$^{1,8}$ \& Inger J\o rgensen$^9$\vspace{4mm}}

\affiliation{$^1$International Centre for Radio Astronomy Research, Uni.~of Western Australia, 7 Fairway, Crawley, WA 6009, Australia;
$^2$Centre for Astrophysics \& Supercomputing, Swinburne Uni.~of Technology, PO Box 218, Hawthorn, VIC 3122, Australia;
$^3$Department of Astronomy and Astrophysics, Uni.~of Toronto, 50 St George St, Toronto, ON M5S3H4, Canada;
$^4$Max Planck Institut f\"ur extraterrestrische Physik, Postfach 1312, Giessenbachstr., D-85741 Garching, Germany;
$^5$Australian Astronomical Observatory, PO Box 915, North Ryde, NSW 1670, Australia;
$^6$Research School of Astronomy and Astrophysics, Australian National Uni., Cotter Rd, Weston, ACT 2611, Australia;
$^7$Harvard-Smithsonian CfA, 60 Garden St., MS-20, Cambridge, MA 02138, USA;
$^8$ARC Centre of Excellence for All-sky Astrophysics (CAASTRO);
$^9$Gemini Observatory, 670 N. A'ohoku Pl., Hilo, HI 96720, USA}

\begin{abstract}
We measure the stellar specific angular momentum $\js=\Js/\Ms$ in four nearby ($z\approx0.1$) disk galaxies that have stellar masses $\Ms$ near the break $\Mknee$ of the galaxy mass function, but look like typical star-forming disks at $z\approx2$ in terms of their low stability ($Q\approx1$), clumpiness, high ionized gas dispersion (40$-$50 $\kms$), high molecular gas fraction (20$-$30\%) and rapid star formation ($\sim\!20~\msunyr$). Combining high-resolution (Keck-OSIRIS) and large-radius (Gemini-GMOS) spectroscopic maps, only available at low $z$, we discover that these targets have $\sim\!3$ times less stellar angular momentum than typical local spiral galaxies of equal stellar mass and bulge fraction. Theoretical considerations show that this deficiency in angular momentum is the main cause of their low stability, while the high gas fraction plays a complementary role. Interestingly, the low $\js$ values of our targets are similar to those expected in the $\Mknee$-population at higher $z$ from the approximate theoretical scaling $\js\propto(1+z)^{-1/2}$ at fixed $\Ms$. This suggests that a change in angular momentum, driven by cosmic expansion, is the main cause for the remarkable difference between clumpy $\Mknee$-disks at high $z$ (which likely evolve into early-type galaxies) and mass-matched local spirals.
\end{abstract}

\maketitle


\section{Introduction and Motivation}\label{section_introduction}

In studying the history of the universe, the galaxy population that contributes most of the star formation density at a given epoch is of central interest. Individual galaxies can evolve into and out of this population, but remarkably the dominant star-forming population itself always remains composed of disks with similar stellar masses of $\Ms=10^{10}$$-$$10^{11}\msun$, at redshifts $z<3$ \citep{Karim2011}. This mass range coincides with the `knee' in the mass function of star-forming galaxies, centered at $\Mknee=10^{10.7}$$-$$10^{11}\,\msun$ at $z<3$ \citep{Muzzin2013}. 

Although the dominant star-forming population always consists of $\Mknee$-galaxies of similar mass, the nature of these galaxies changes significantly with $z$. In modern times ($z\approx0$), most of these galaxies are stable spiral disks like the Milky Way. They have low velocity dispersion and balance gravity almost entirely by rotation. Stars form at modest rates ($\sim\!1\,\msunyr$) in thousands of clouds with individual masses around $10^5$$-$$10^6\,\msun$ \citep{Murray2011}. In contrast, galaxies of the same mass in past epochs ($z\approx1$$-$$3$) form stars 10--100 times more rapidly. Most of this fast star formation occurs in giant gas clumps of $10^8$$-$$10^9\,\msun$ \citep{Elmegreen2005}, reminiscent of merging proto-galaxies, but now interpreted as large Jeans instabilities \citep{Bournaud2007}. These disks are rich in turbulent gas \citep{Genzel2008} and yield marginal rotational support and stability \citep{ForsterSchreiber2006}. Because of mass growth, these clumpy $\Mknee$-disks at $z\approx2$ are likely to evolve into early-type galaxies (ETGs, \citealp{Tacchella2015}).

The stark morpho-dynamical difference between early and modern star-forming galaxies, in spite of identical masses, raises the question as to how these galaxies `know' what cosmic time they are at. Is their difference explained by a change in the mass composition, such as the established decline in the cold gas fraction\citex{Tacconi2010} and the molecular-to-atomic ratio\citex{Obreschkow2009c}, or are these paralleled by a more fundamental driver? Angular momentum is a promising contender because of its predicted $z$-dependence at fixed galaxy mass: the stellar specific angular momentum $\js\equiv\Js/\Ms$ is expected to scale as  $\js\propto(1+z)^{-1/2}$ for fixed stellar mass $\Ms$ (not individual galaxies) as discussed in Section~\ref{subsection_discussion_lowz}. This motivates the question whether low angular momentum drives or contributes to the marginal stability of the clumpy, turbulent $\Mknee$-disks commonly found at high $z$. This idea is further stimulated by the empirical fact that $\js$ is also the dominant driver of morphological differences in regular galaxies of equal mass at $z\approx0$ \citep{Romanowsky2012,Obreschkow2014a}. 

Direct angular momentum measurements of clumpy, turbulent disks at $z>1$ are, however, challenged by difficulties of conducting deep, high-resolution spectroscopic observations at such large cosmic distances. Overcoming this challenge was the motivation to assemble the `DYNAMO' galaxies\citex{Green2014}, a sample of 95 galaxies containing some of the most H$\alpha$ luminous ($L_{\rm H\alpha}\!>\!10^{42}\rm erg\,s^{-1}$) nearby ($z\approx0.1$) galaxies, excluding AGN, in the Sloan Digital Sky Survey\citex{York2000}. Follow-up observations (Section \ref{section_data}) revealed that, merging objects aside, the H$\alpha$ luminous DYNAMO objects are analogous to typical clumpy disks at $z\approx1-3$ (details in Section~\ref{section_data}). Hence, the DYNAMO sample is an interesting laboratory to probe the angular momenta in clumpy galaxies, normally too distant for such measurements.

The objective of this article is to answer two questions: (i) what role does angular momentum play in shaping the rare clumpy, turbulent disks at low-$z$, and (ii) what can these low-$z$ studies tell us about the high-$z$ star-forming population? These questions will be examined using a pilot sample of four clumpy DYNAMO disks with exquisite integral field spectroscopy (IFS) data. Section \ref{section_data} describes the sample and summarizes the kinematic measurements, expanded in the Appendix. Section \ref{section_discussion} discusses and interprets the results, synthesized in Section \ref{section_conclusions}.

\section{Sample and Analysis}\label{section_data}

We perform the angular momentum measurements in a pilot sample of four DYNAMO galaxies at $z\approx0.1$ (Figure~1 and Table~1), which exhibit H$\alpha$ luminosities in the top $0.1\%$ at their redshifts, but have regular rotation and do \emph{not} show the typical merger/starburst signatures of Ultra Luminous Infrared Galaxies. The targets have stellar masses near $\Mknee$, similar to the Milky Way ($\log(\Ms/\msun)=10.8$, \citealp{McMillan2011}), but appear to be highly reminiscent of typical star-forming galaxies at $z\approx2$ given the following properties: (i) they are composed of giant star-forming clumps revealed in HST H$\alpha$-imaging \citep{Bassett2014}, \new{(ii) they exhibit high star formation rates (mean of $20~\msunyr$) comparable to main-sequence galaxies at $z\geq1$ \citep{Green2014}, (iii) they have turbulent (flux weighted gas dispersion $\sigma=40$$-$$50~\kms$), but rotationally supported disks ($V/\sigma=3$--$5$) placed on a Tully-Fisher relation (\citeauthor{Green2014}), (iv) the disk sizes ($\re\approx3$~kpc) are comparable to those of $\Mknee$-disks at $z\geq1$ (\citeauthor{Green2014})}, (v) CO(1--0) emission data uncovers high molecular gas fractions (20$-$30\%, \citealp{Fisher2014}).

Given the strong analogy to high-$z$ turbulent disk galaxies, our objects represent an exquisite laboratory for studying such clumpy disks at distances 20-times closer than possible so far. \new{To our knowledge, the DYNAMO sample is unique in that it fulfils \emph{all} the above properties, including $\Ms\approx\Mknee$. Therefore, our sample is a useful addition to the well-established class of `low-z analogs of high-z objects', reviewed by \cite{Glazebrook2013} and \cite{Elmegreen2013}.}

Our measurements of $\js$ (details in Appendix A \& B) rely on high-resolution adaptive optics IFS data from Keck-OSIRIS to probe the inner parts and, in two objects (G04-1 and G20-2), natural seeing Gemini-GMOS IFS to probe the faint outer parts to 2.5$R_e$. These data give, in each pixel, stellar surface densities from the broad-band infrared/optical continuum and line-of-sight velocities from Doppler-shifted hydrogen emission lines (Pa$\alpha$ and H$\beta$). \new{In the case of G04-1 and G20-2, we were also able to extract stellar absorption line maps, revealing that the velocities \emph{and} dispersions of the stars are consistent with those of the ionized gas \citep{Bassett2014}. We can therefore adopt a single rotation curve for gas and stars}. Where available, data from both spectrographs is combined (Appendix~A). We note that the OSIRIS data confirms the high velocity dispersion found by \cite{Green2014} (45$-$50~$\kms$ for G04-1, G20-2) at a high spatial resolution where the effect of beam smearing is negligible (Oliva et al. in prep.).

Kinematic models are fitted \textit{simultaneously} to the combined continuum-density and emission line-velocity data, making no assumption on the rotation structure other than circular rotation in a disk (Appendix~B). The fitting routine was successfully verified against degraded mock data of 1,000 model-galaxies and the control galaxy NGC~3198 (see Appendix~C).

The fits to all targets (Figure~1) are excellent in that no global (\mbox{mono-,} \mbox{di-,} or quadrupole) systematics can be seen in the density and velocity residuals. Table~1 lists the fitting parameters and two values for $\js$, one ($\jobs$) based on the pixels that have IFS data, and one ($\jextra$) that also includes an extrapolation to infinite radii assuming an asymptotically flat rotation and exponential disk for the outer regions without data. For the two galaxies with only inner rotation data, to one effective radius $\re$, from Keck-OSIRIS, the fraction of the total specific angular momentum $\jextra$ that is extrapolated is about 60\%, and the estimated statistical uncertainty of $\jextra$ amounts to about 18\%. In the two galaxies with additional natural seeing Gemini-GMOS data, reaching beyond $2\,\re$, only about 25\% of $\jextra$ is extrapolated and hence the uncertainty of $\jextra$ reduces to about 11\%. This shows the utility of such deep IFS data for measuring angular momentum.

In Figure~2, we compare the angular momenta of the clumpy targets to typical spiral galaxies at $z\approx0$ in the mass-spin-morphology space, spanned by $\log\Ms$, $\log\js$, and the stellar bulge-to-total mass ratio $B/T$. The regular spiral galaxies of The HI Nearby Galaxy Survey (THINGS, \citealp{Walter2008}) lie near a plane in this space\citex{Obreschkow2014a}. In these data, the bulge masses in the $B/T$ ratios were defined as the central excess above an exponential density profile (Appendix~A of \citeauthor{Obreschkow2014a}), irrespective of the bulge types (mostly pseudo-bulges). The $B/T$ values of our four targets were computed in the same way from radial density profiles, extracted from HST $r$-band (filter FR647M) images (for D13-5, G04-1, G20-2) and OSIRIS $K$-band continuum maps (for C22-2).

For consistency, all stellar masses in Figure~2 were estimated from $K$-band magnitudes\footnote{$K$-magnitudes (Vega zero-point) are from 2MASS for DYNAMO and from $3.6\mu\rm m$ Spitzer maps with a 2MASS-calibrated $3.6\mu\rm m$-to-$K$ conversion for THINGS\citex{Leroy2008}.} within infinite apertures\footnote{For DYNAMO objects, the 2MASS aperture is $\sim\!4$ scale radii containing 99\% of the light. For THINGS, the stellar masses in\citex{Leroy2008} were calculated using apertures of 1.5$r_{25}$. By extending the aperture to $\gtrsim5\re$, we find the flux to increase by a factor 1.2. This factor was applied to all THINGS galaxies.}, using a small empirical $k$-correction\citex{Glazebrook1995} and a universal $K$-band mass-to-light ratio\footnote{Assuing a \cite{Kroupa2001} Initial Mass Function (IMF) for consistency with the THINGS data \citep{Leroy2008}.} $M/L_{\rm K}=0.5\,\msun/\lsun$. Using a color-dependent $M/L_{\rm K}$ is more adequate in studies including galaxies with significant classical bulges that tend to be redder \citep{Fall2013}. However, in the present comparison a constant $M/L_{\rm K}$ suffices: the THINGS and DYNAMO galaxies are disk-dominated and their continuum maps are consistent with pseudo-bulges \citep{Kormendy2008}, except for three THINGS objects with small classical bulges. Overall, the DYNAMO galaxies tend to be bluer than typical spirals, but the $M/L_{\rm K}$ ratio is insensitive (to $<20\%$) to this\citex{Bell2003}.

\begin{table*}
	\footnotesize
	{\centering\textbf{Table 1}\\}
	{\centering Properties of the four clumpy target galaxies and the control galaxy NGC~3198.\vspace{1mm}\\}
	\renewcommand{\arraystretch}{1.4}
	\noindent \begin{tabular}{l|ccc|cc|cccccccc}
	\hline\hline
	& \multicolumn{3}{c|}{Basic properties} & \multicolumn{2}{c|}{IFS Resolution$^c$} & \multicolumn{8}{c}{Fitted in this work$^d$} \\
	Object~~~~~~ & ~Redshift~ & $\Ms^a$ & $L_{\rm H\alpha}^b$ & OSIRIS & GMOS & $\incl$ & $\alpha$ & $B/T$ & $\re$ & $r_{\rm flat}$ &  $v_{\rm flat}$ & $\!\jobs\!$ & $\!\!\!\!\jextra\!\!\!\!$ \\
	& & $\log(\msun)$ & $\log(\rm erg\,s^{\!-1})$ & kpc & kpc & deg & deg & & kpc & kpc & $\!\rm km\,s^{\!-1}\!$ & $\!\rm kpc\,km\,s^{\!-1}\!$ & $\!\rm kpc\,km\,s^{\!-1}\!$ \\
	\hline
	C22-2&0.07116&$\!10.4\!$&$\!41.68\!$&$\!0.33\!$&--&50&145&$\!\!\!0.12\!\!\!$&$\!2.6\!$&--&--&120&$273\pm49$\\
D13-5&0.07535&$\!10.7\!$&$\!42.10\!$&$\!0.35\!$&--&38&127&$\!\!\!0.01\!\!\!$&$\!3.5\!$&0.5&202&356&$828\pm143$\\
G04-1&0.12981&$\!11.1\!$&$\!42.36\!$&$\!0.57\!$&$\!1.19\!$&40&359&$\!\!\!0.15\!\!\!$&$\!5.1\!$&0.5&135&515&$834\pm116$\\
G20-2&0.14113&$\!10.7\!$&$\!42.26\!$&$\!0.62\!$&$\!1.28\!$&46&187&$\!\!\!0.18\!\!\!$&$\!2.7\!$&0.6&118&337&$394\pm32$\\
\hline
NGC 3198&$\!\!662\,\kms\!\!\!$&$\!10.2\!$&$\!\!$&$\!0.02\!$&--&69&215&$\!\!\!0.00\!\!\!$&$\!6.2\!$&3.9&163&626&$973\pm122$\\

	(reference) & & & & & & $\!\!\!\!72^e\!\!\!\!$ & $\!\!\!\!215^e\!\!\!\!$ & $\!\!\!\!0.03^f\!\!\!\!$ & & $\!\!\!\!2.8^e\!\!\!\!$ & $\!\!\!\!150^e\!\!\!\!$ & & $\!\!\!\!936\pm20$$^f\!\!\!\!$ \\
	\hline
	\hline
	\multicolumn{13}{c}{}\vspace{-2mm}
\end{tabular}\\
	{\textbf{Notes.} $^a$\,$K$-band based stellar masses ($\Upsilon_K=0.5\,\msun/\lsun$); $^b$\,H$\alpha$ luminosity as measured by IFS observations on the AAT/SPIRAL\citex{Green2014}; $^c$\,FWHM of PSF; $^d$\,Parameters fitted as described in Appendix~B; $^e$\,Table~4 of \cite{Leroy2008}; $^f$\,Table~1 of \cite{Obreschkow2014a}.}
\end{table*}

\begin{figure*}
	\includegraphics[width=\textwidth]{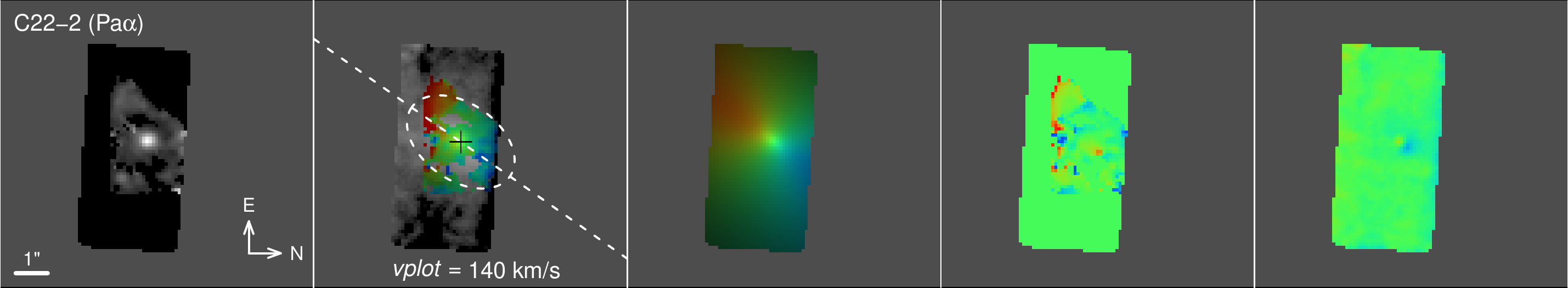}
	\includegraphics[width=\textwidth]{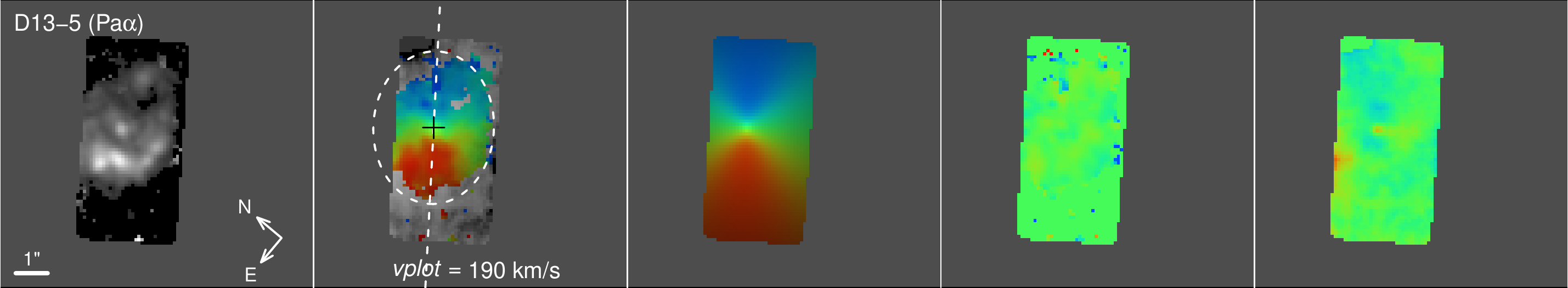}
	\includegraphics[width=\textwidth]{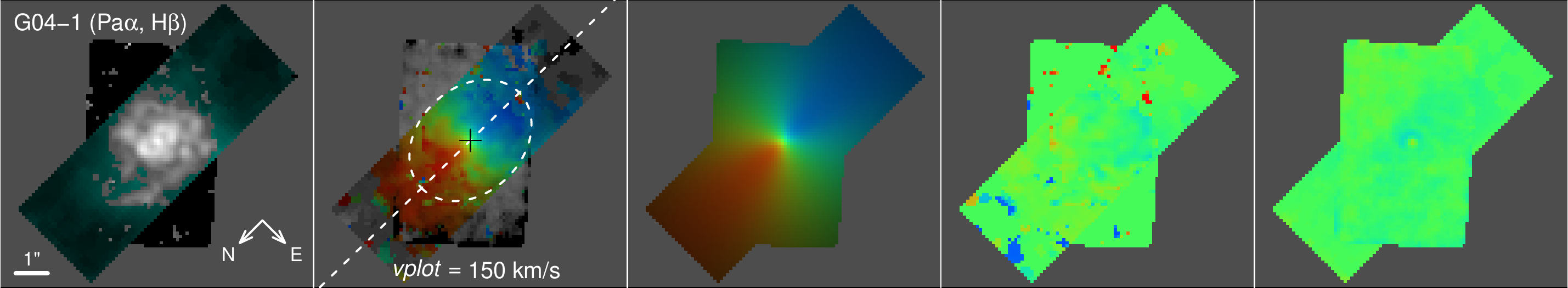}
	\includegraphics[width=\textwidth]{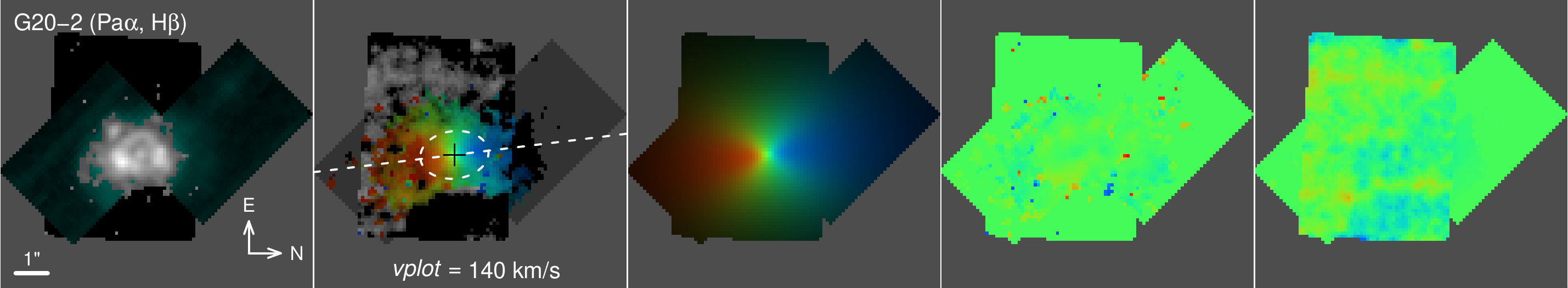}
	\includegraphics[width=\textwidth]{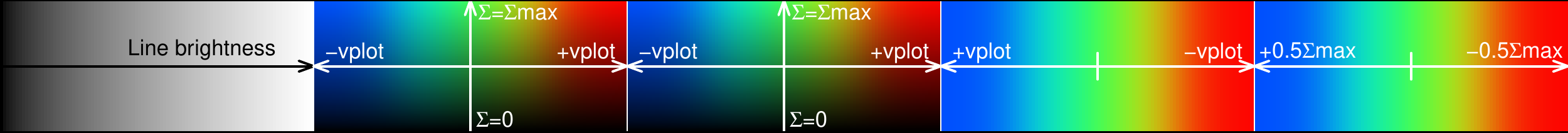}\vspace{2mm}
	{\textbf{Figure 1.} Kinematic fits for the four clumpy targets. Col 1: emission line fluxes of the Pa$\alpha$ (OSIRIS) and H$\beta$ (GMOS) lines, as indicated in parentheses; H$\beta$ is shown in cyan for distinction from Pa$\alpha$. Cols 2\&3: density+velocity maps of data and models, respectively. Lightness represents the continuum surface brightness and hue represents the rest-frame line-of-sight velocity of the emission lines. A greyscale is used where only continuum data is available. White dashed lines show the best-fit major axes and half-mass ellipses. Cols 4\&5: Maps of the line-of-sight velocity residuals $\Delta v'=v'-v'_{\rm model}$ and stellar density residuals $\Delta\Ss=\Ss-\Sigma_{\rm model,s}$.}
\end{figure*}

\section{Results and Discussion}\label{section_discussion}

\subsection{Low $\js$ in clumpy galaxies at $z\approx0$}\label{subsection_discussion_lowz}

Figure~2 reveals a clear offset of the DYNAMO targets from the THINGS galaxies, corresponding to a factor $3.0\pm0.7$ shift in $\js$: the clumpy targets have three times less angular momentum than regular local spiral galaxies of comparable $\Ms$ and $B/T$. A likelihood analysis (using the \textit{Hyper-Fit} package of \citealp{Robotham2015}), accounting for the statistical uncertainties and small number statistics, shows that the offset between the two samples is certain at 5.1 standard deviations and larger than a factor 2 in $\js$ with 97\% probability.

The immediate suggestion of Figure~2 is that the difference between turbulent, clumpy disks and regular spiral disks is in some sense related to angular momentum. The question then becomes one of causality: is the \emph{leading} mechanism that (i) low angular momentum causes clumps, (ii) clumps cause low angular momentum, or (iii) a third quantity causes both? Options (ii) and (iii) are hard to substantiate in light of angular momentum being a conserved quantity that cannot be generated from within the galaxy-halo system. Thus, option (iii) would mean that external torques somehow cause the giant clumps while also reducing the angular momentum -- an implausible scenario given that our targets show no global kinematic asymmetries hinting at such torques. For option (ii) to work, the star-forming clumps would need to drive a mechanism (e.g.~via tidal torques or ejected material) for transferring angular momentum from the disk to the dark halo. However, in modern high-resolution simulations\citex{Genel2015} star-forming regions preferentially remove low-$j$ gas via supernova-winds, increasing the angular momentum of the disk rather than reducing it, in order to make simulated disk galaxies as large as observed. In contrast, option (i), i.e. low-$j$ causing clumps, appears plausible from the following fundamental stability considerations.

The stability of a galactic disk against clump formation can be quantified by the dimensionless Toomre parameter \citep{Toomre1964}\footnote{\new{In the case of pure gas disks $\sigma_{\rm R}$ is commonly expressed via the sound-speed $c_{\rm s}$, which is well-approximated by the dispersion.}}
\new{\be\label{eq_toomre_def}
	Q = \frac{\sigma\kappa}{3.36\,G\Sigma},
\ee
which depends on the local surface density $\Sigma$, radial velocity dispersion $\sigma$, and epicyclic frequency $\kappa$}. If $Q<1$, instabilities can develop with diameters larger than the Jeans length $L_{\rm J}$ and smaller than the maximal stability scale $L_{\rm R}$ set by the differential rotation. If $Q>1$, then $L_{\rm J}>L_{\rm R}$ and no instability can develop: all Jeans modes are stabilised by the local centrifugal acceleration.

In most galactic disks, the surface density and epicyclic frequency decrease similarly with radius, such that $Q$ is constant within $20\%$ (for individual galaxies, \citealp{Westfall2014}) across the regions that hold most of the stellar mass (at $\sim1$ exponential radius) and angular momentum (at $\sim2$ exponential radii). Within this level of accuracy it is therefore possible to introduce a mean Toomre parameter $\Qm$, defined such that $\Qm>1$ corresponds to globally stable disks.  \new{The global stability parameter $\Qm$ can be rewritten as $\Qm\propto M^{-1}j\sigma$ upon expressing the means of $\Sigma$ and $\kappa$ by the total mass $M$ and specific angular momentum $j$ of the disk \citep[details in][]{Obreschkow2014a}. This straightforward substitution does \emph{not} change the underlying physics, but introduces a new viewpoint, describing the disk stability in terms of fundamental quantities, conserved in isolation. As we shall see, this viewpoint allows us to link the disk stability to a time-dependent cosmological context.}

\new{So far, $\Qm$ describes the stability of a single component (stellar or gas) disk. To about 20\% accuracy \citep{Romeo2011} this quantity can be extended to a multi-component (stars and gas) disk via \citep{Wang1994}
\be
	\Qm^{-1}=\Qm_{s}^{-1}+\Qm_{\rm g}^{-1},
\ee
where $\Qm_{s}\propto M_{s}^{-1}j_{s}\sigma_{s}$ and $\Qm_{g}\propto M_{g}^{-1}j_{g}\sigma_{g}$ are the partial stability terms of the stars and cold gas, respectively. Relying on the observational finding (see \S2 and \citealp{Bassett2014}) that stars and gas exhibit comparable velocity and dispersion maps, we here adopt the equalities $\js=\jg\equiv j$ and $\sigma_s=\sigma_{\rm g}\equiv\sigma$. This does not necessarily imply that the gas and stars develop similar instabilities, as discussed at the end of \S\ref{subsection_discussion_lowz}.}

Introducing the gas fraction $\fg\equiv\Mg/(\Ms+\Mg)$, the global Toomre parameter then reduces to
\be\label{eq_toomre}
	\Qm \propto \Ms^{-1}(1-\fg)\,\js\,\sigma.
\ee
Main-sequence spiral galaxies in the local universe are stable systems with $\Qm\approx\Qm_{\rm local}=2$ \citep{Westfall2014}.

\new{We now discuss how $\Qm$ changes relative to $\Qm_{\rm local}$ in the case of our clumpy turbulent disks.} \eq{toomre} shows that the high velocity dispersion $\sigma$ of our targets, which has about twice the amplitude of typical local disks \citep{Andersen2006}, would by itself \emph{increase} the stability by a factor 2. The reason is that $L_{\rm J}$ increases with $\sigma$, a situation that is also observed in many high-$z$ disks \citep{Genzel2008}. Thus, the high $\sigma$ can explain the large Jeans scale (Fisher \textit{et al.}, forthcoming), but it cannot simultaneously explain why the stability of the disks is low enough (i.e.\ $L_{\rm R}$ is high enough) for the Jeans modes to become manifest. Following \eq{toomre}, this low stability despite high $\sigma$ requires high gas fractions or low angular momentum (at fixed $\Ms$). Our targets have both: $\fg=20$$-$30\% (seen in CO emission\footnote{The 20$-$30\% refer to molecular material, observed in CO emission\citex{Fisher2014}. An additional atomic (\ha) component could increase these percentages, but is expected to be situated at larger radii, negligible for the stability of the stellar component, in such dense disks\citex{Obreschkow2009c}.}) is a factor 2--3 above the typical 10\% value of local $\Mknee$-disks \citep{Leroy2008}, while the $\js$ values measured in this work are a factor \mbox{$\sim\!3$} lower than in typical local $\Mknee$-disks.

Importantly, \eq{toomre} shows that the factor $1/3$ reduction in $\js$ has a much larger direct effect on reducing $\Qm$ than the factor 2--3 increase in $\fg$. Thus, \emph{low angular momentum is the dominant driver for the low stability of our targets} -- the main result of this study. Quantitatively, the $\Qm$-values of our targets are about half those of typical local spirals, due to the combination of a factor $0.8$ decrease in $(1-\fg)$, a factor $0.3$ decrease in $\js$, and a factor 2 increase in $\sigma$, resulting in $\Qm=0.8\times0.3\times2\times\Qm_{\rm local}\approx1$. How this low stability feeds back onto $\sigma$, via instability-driven turbulence and stellar feedback, is a topic of current research beyond this study.

\new{It is worth discussing some potential caveats of this stability discussion. Firstly, \eq{toomre_def} relies on a purely mechanical model of a flat disk. In the case of thick disks, the constant 3.36 decreases \citep{Peng2001}, while energy dissipation within clumps can increase this constant. In the present context, where relative changes in $\Qm$ matter more than its absolute value, these effects only appear at second order, unlikely to affect the general results. Secondly, we have omitted the fact that the stellar component is often less clumpy than the gas -- a generic observation in most turbulent disks. So the question arises if this difference in stellar and gas clumpiness can be explained in the context of a single stability parameter $\Qm$. While this question is beyond the scope of this paper, its answer probably requires a more detailed analysis of the dynamics of instabilities. For instance, the evolution of a Jeans instability depends on the adiabatic index $\gamma$ of the fluid: for molecular gas ($\gamma<4/3$), the Jeans mass decreases with increasing density and hence any proto-instability continues its collapse while fragmenting into smaller clumps. In turn, the stellar component behaves roughly like a monoatomic gas ($\gamma=5/3$), in which the Jeans mass increases with increasing density. Hence the collapse is stopped making the stellar disk less clumpy.}

\begin{figure*}
	\hspace{1.0cm}
	\includegraphics[width=13cm]{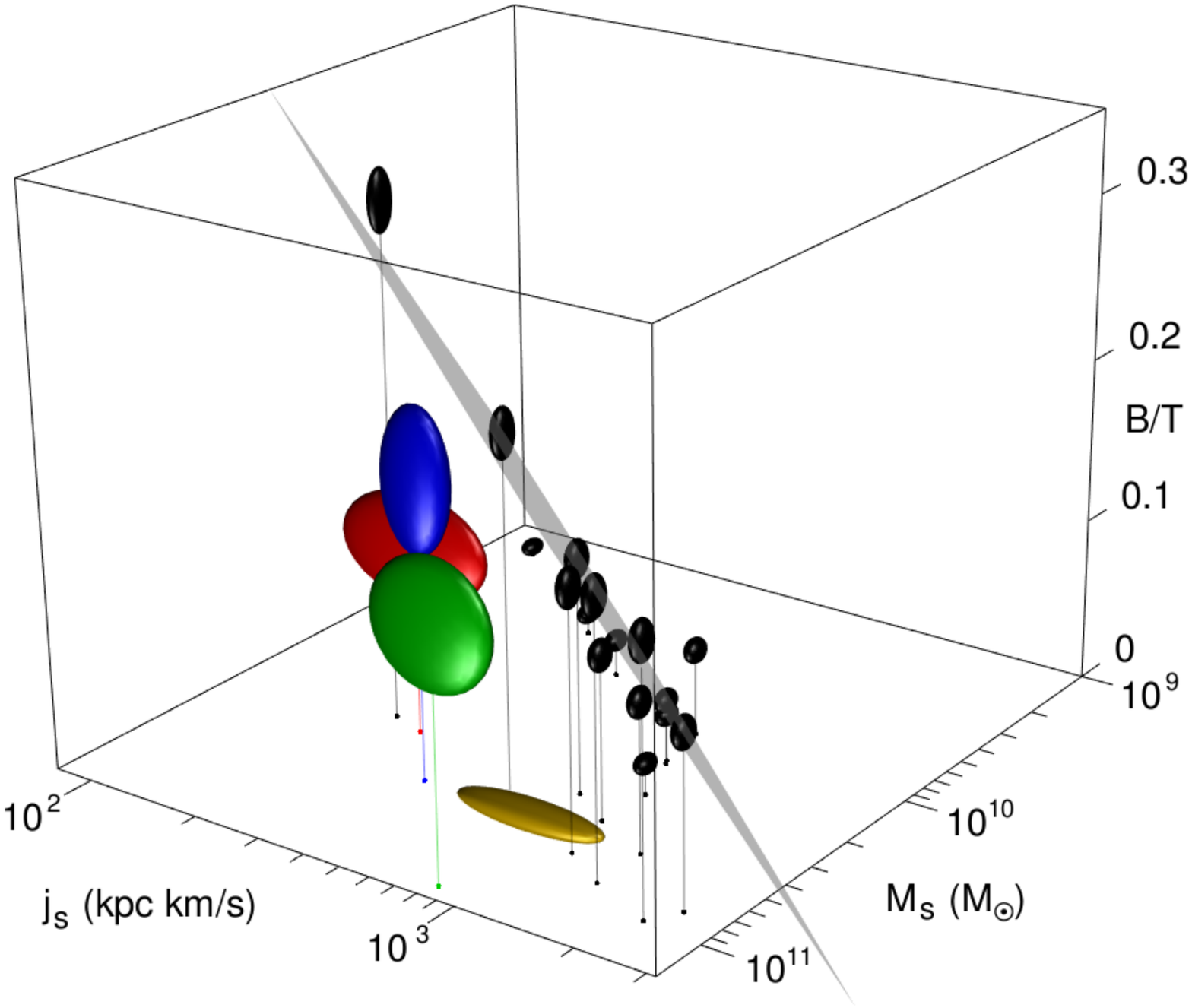}
	\includegraphics[width=3.8cm,trim=1 1 1 1,clip]{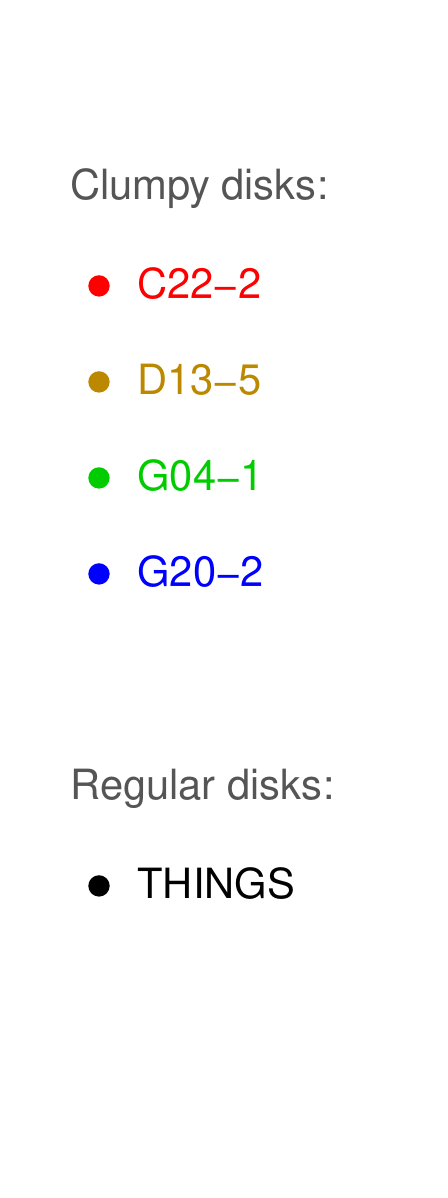}\vspace{-0.5mm}\\
	\textbf{Figure 2.} \href{https://www.dropbox.com/sh/owj3ykdvlcxp3n6/AAC_mhm5_fMqspILoFczxvrWa}{\link{[Interactive 3D-figure in Adobe PDF Viewer. Feel free to use the standalone and animated versions in our Dropbox.]}} Mass-spin-morphology space. Dark points represent the 16 typical local spiral galaxies of the THINGS sample\citex{Obreschkow2014a,Walter2008}; colored points are the four clumpy disks of this study -- they appear to be deficient in angular momentum $\js$. Ellipsoids represent $1\sigma$-statistical uncertainties. The $\js$-values of the DYNAMO objects are the best estimates $\jextra$ (see Appendix~B) listed in Table~1. The grey plane is best fits to the data accounting for the 3D Gaussian uncertainties.\vspace{2mm}
\end{figure*}

\subsection{Connection to the clumpy population at $z\approx1-3$}\label{subsection_discussion_highz}

Let us now turn to our second objective, the extension of the discussion to early cosmic times. While our four targets are exotic at $z\approx0$, their high dispersion, clumpiness, star formation and gas fraction are representative of the main sequence disk population of the same stellar mass at $z\approx1-3$ (Section~\ref{section_data}), now thought to have formed more stars than merger-driven starbursts \citep{Rodighiero2011}. It is clear that the massive star formation activity (10$-$$100\,\msunyr$) of this high-$z$ $\Mknee$-population must be fuelled by copious inflows of fresh gas \citep{Dekel2009}, which in turn feeds from the intergalactic medium that is $\sim\!3^3=27$ times denser at $z\approx2$ than today. However, we still require an explanation for the low disk stability enabling this fresh gas to turn efficiently into molecules and stars \citep{Tacconi2010}. Is the gas itself the destabilizing agent, or do these high-$z$ $\Mknee$-disks exhibit low angular momentum, just as we discovered in our four local high-$z$ analogs?

Measuring the angular momentum in disks at $z\approx2$ is extremely difficult, as their inner rotation structure ($<0\farcs1$) is hard to resolve and their outer parts, holding most of the angular momentum, have low surface brightness (owing to the \mbox{$\sim(1+z)^{-4}$ cosmological dimming)}.
However, a generic theoretical expectation in \lcdm cosmology is that the mean relation between mass and angular momentum changes with time, due to cosmic expansion. To first order, this can be seen in the model of a spherical halo (virial mass $\Mh$, radius $\Rh$, velocity $\Vh$, angular momentum $\Jh$). Following \cite{Mo1998}, the specific angular momentum $\jh\equiv\Jh/\Mh=\sqrt{2}\lambda RV$ can be expressed as
\be\label{eq_jz}
	\jh=2^{2/3}G^{2/3}\lambda\Mh^{2/3}H(z)^{-1/3}\Delta_{\rm c}(z)^{-1/6},
\ee
where $G$ is the gravitational constant. The spin parameter $\lambda$, here defined as in \citealp{Bullock2001}, is approximately independent of $z$ \citep{Hetznecker2006}. The Hubble parameter varies as $H(z)=H_0 E(z)$, where $H_0$ is the local Hubble constant and $E(z)=(\Omega_{\rm m}(1+z)^3+\Omega_\Lambda)^{1/2}$ in a flat universe. The function $\Delta_{\rm c}(z)$ is the over-density factor of the virialized halo relative to the mean density of the universe, approximated analytically as \citep{Bryan1998}
\be\label{eq_Dz}
	\Delta_{\rm c}(z) = 18\pi^2-82\Omega_\Lambda E^{-2}(z)-39\Omega_\Lambda^2E^{-4}(z).
\ee
Substituting these analytical expressions for $H(z)$ and $\Delta_{\rm c}(z)$ into \eq{jz} yields $\smash{\jh\propto\Mh^{2/3}(1+z)^{-1/2}}$, where the $z$-dependance is an approximation. Assuming a $z$-independent $\Ms/\Mh$ (as verified to $z\lesssim4$, \citealp{Behroozi2013}) and retention of the baryon specific angular momentum during disk formation\citex{Fall1980}, a similar $z$-scaling applies to $\js$,
\be\label{eq_js}
	\js\propto\Ms^{2/3}(1+z)^{-1/2}.
\ee

Modern hydrodynamic galaxy simulations in a cosmological context, the Magneticum Pathfinder \citep{Hirschmann2014} and Illustris \citep{Vogelsberger2014} simulations, produce robust angular momentum statistics \citep{Genel2015,Teklu2015}. Both approximately confirm the $z$-scaling of \eq{js} for star-forming $\Mknee$-disks (priv.\ comm.\ S.\ Genel and A.\ Teklu).

In summary, there is strong theoretical indication that the population of $\Mknee$-disks at $z\approx2$ has lower $\js$ (by a factor $\sim\!2$) than the mass-matched disk population at $z\approx0$. The expected $\js$ values of the high-$z$ disks are consistent with to those measured in our local analogs within the uncertainties (at 1.3 standard deviations). Following the stability discussion of Section~\ref{subsection_discussion_lowz}, low angular momentum must then play a major role in making the high-$z$ star-forming disks semi-stable. It thus seems that the cosmic growth of angular momentum is a dominant driver behind the evolution of the population of star-forming $\Mknee$-galaxies from semi-stable, turbulent, clumpy to stable, flat, regular spiral disks. We emphasize that this conclusion applies to the evolution of the \emph{population} of $\Mknee$-galaxies, not individual galaxies. A $\Mknee$-galaxy at $z\approx2$ does not normally evolve into a $\Mknee$-galaxy at $z\approx0$, but into a much more massive object \citep{VanDokkum2013} such as an ETG; nor are individual disks expected to follow $\js\propto(1+z)^{-1/2}$.

A potential caveat is that typical $z\approx2$ disks of stellar mass near $\Mknee$ may have even higher gas fractions ($\sim\!50\%$, \citealp{Carilli2013}) than our objects, which would contribute an extra stability factor of 40\%. However, this is still sub-dominant to the change in $j$, and may arise from it.

\section{Conclusions}\label{section_conclusions}

This article reports the detection of exceptionally low angular momentum in four rare examples of clumpy, turbulent star-forming galaxies at $z\approx0.1$. The low angular momenta offer a powerful new explanation for their marginal stability, enabling large clumps to form and pristine gas to convert rapidly into stars. The key to obtaining robust measurements of specific angular momentum is the combination of IFS data with high spatial resolution in the galaxy centers (from Keck-OSIRIS) and high surface brightness sensitivity out to large radii (from Gemini-GMOS).

The circumstances through which clumpy low-$j$ disks with high gas fractions can exist in the local Universe remain to be explored, but whatever the origin of these objects, it is likely that the physical mechanisms driving clump formation and turbulence are the same as at $z\approx2$ and independent of the galaxies' origin.  By exploiting the analogy between our targets and \emph{typical} star-forming galaxies at $z\approx2$, we find evidence that low angular momentum is also a dominant cause of the low stability of such high-$z$ galaxies. Hence, a change in angular momentum, driven by cosmic expansion, appears to be the primary cause for the remarkable difference between clumpy $\Mknee$-disks at high $z$ (which likely evolve into ETGs) and mass-matched local spirals.\\
\\

{\footnotesize \emph{We dedicate this publication to the memory Prof Peter McGregor, who passed away during this work. We are indebted to him for his outstanding contributions to astrophysics, his leadership of Integral Field Spectroscopy in Australia and his inspiration to the DYNAMO project.} \\ \\ Acknowledgements: Based on observations made with the NASA/ESA Hubble Space Telescope, obtained at the Space Telescope Science Institute, which is operated by the Association of Universities for Research in Astronomy, Inc., under NASA contract NAS 5-26555. These observations are associated with program \#12977. --- We acknowledge observational support from the Swinburne Keck Time Assignment Committee, and financial support from ARC Discovery Project DP130101460. This work is also based on observations obtained at the Gemini Observatory (programs GN-2011B-Q-54 and GS-2011B-Q-88), which is operated by the Association of Universities for Research in Astronomy, Inc., under a cooperative agreement with the NSF on behalf of the Gemini partnership: the National Science Foundation (United States), the National Research Council (Canada), CONICYT (Chile), the Australian Research Council (Australia), Minist\'erio da Ci\^encia, Tecnologia e Inova\c{c}\~ao (Brazil) and Ministerio de Ciencia, Tecnolog\'ia e Innovaci\'on Productiva (Argentina). R.B. acknowledges support from the Victorian Department of State Development, Business and Innovation through the Victorian International Research Scholarship (VIRS). --- Some of the data presented herein were obtained at the W.M. Keck Observatory, which is operated as a scientific partnership among the California Institute of Technology, the University of California and NASA. The Observatory was made possible by the generous financial support of the W.M. Keck Foundation. The authors wish to recognize and acknowledge the very significant cultural role and reverence that the summit of Mauna Kea has always had within the indigenous Hawaiian community.\vfill}

\appendix

\subsection{A. IFS data and pre-processing}

We observed all four targets with the Keck-OSIRIS integral field spectrograph in Adaptive Optics (AO) mode to obtain spectra around the Pa$\alpha$ (1875\,nm rest-frame) line in $5\farcs2\times6\farcs7$ fields. The spatial resolution is $0\farcs24$ full-width-half-maximum (FWHM) of the point spread function (PSF), corresponding to the sub-kpc resolutions listed in Table~1. For the two more distant galaxies (G04-1 and G20-2) we also obtained deeper IFS data around the H$\beta$ (486.1\,nm rest-frame) line using the Gemini-GMOS at a natural seeing of $0\farcs5$ in three fields (one for G04-1, two for G20-2, visible as 45$^\circ$ rotated rectangles in Figure~1).

The 3D-data cubes (two spatial dimensions, one wavelength dimension) from OSIRIS and GMOS were converted into rest-frame line-of-sight velocity maps $v'(\x)$ (from Pa$\alpha$, H$\beta$) and background-subtracted continuum flux maps $\Ss(\x)$, on a discrete cartesian grid $\x_i=(x_i,y_i)$ with $0\farcs1\times0\farcs1$ pixels. We assume the stellar surface density to be proportional to $\Ss(\x)$ (hence the subscript).
We combine the OSIRIS and GMOS data (for G04-1 and G20-2) by registering the maps at $0\farcs1$-pixel accuracy, which suffices given the $0\farcs5$ PSF of GMOS. The GMOS continuum values are scaled to match the OSIRIS continuum at the edge of the OSIRIS maps. Merged continuum and velocity maps are then created by taking OSIRIS data where available and GMOS data otherwise. In this way, the central parts, where spatial resolution is critical, always come from OSIRIS.

\subsection{B. Kinematic fits and angular momentum measurement}

The surface density maps $\Ss(\x)$ and line-of-sight velocity maps $v'(\x)$ on a discrete grid of $0\farcs1$-pixels $\x_i=(x_i,y_i)$ are fitted with a disk model using a customised algorithm:
\begin{enumerate}
\itemsep0.2em
\item Fit galaxy centre $\x_0=(x_0,y_0)$ by minimizing the convolution of the density-weighted velocity map $w=\Ss v'$ and its mirror image, $\int\!d^2x~w(\x-\x_0)w(\x_0-\x)$. This method is exact for axially symmetric objects.
\item Fit inclination $\incl$ and position angle $\alpha$, which set the projection factor $f_{\incl,\alpha}(\x)=(\cos^2\!\varphi+\sin^2\varphi\cos^{-2}\incl)^{1/2}$, $\varphi=\arg(\x-\x_0)-\alpha$, such that the radius in the galaxy plane is $r=f_{\incl,\alpha}(\x)|\x\!-\!\x_0|$, and the factor $g_{\incl,\alpha}(\x)=f_{\incl,\alpha}(\x)(\cos\varphi\sin\incl)^{-1}$, such that the circular velocity is $v(r)=g_{\incl,\alpha}(\x)v'(\x)$. The parameters $\incl$ and $\alpha$ are fitted simultaneously without assuming a rotation or density structure other than a flat, axial disk. This fit is performed by minimizing the RMS of the uncertainty-weighted $\Ss$ and $v$ residuals in circular rings in the galaxy plane. We do not a account for beam-smearing in the velocity field, since the central steep parts of the rotation curve are sampled with adaptive optics in all four targets. Explicitly, the maximal relative velocity change per PSF FWHM $\Delta x$, calculated as $\max\{\nabla v'(\x)\}\Delta x~v_{\rm flat}^{-1}$, is much smaller than unity in all cases.
\item Fit a four-parameter disk+bulge density profile $\Sigma_{\rm model,s}(r)=\Sigma_{\rm d}\exp(-r/r_{\rm d})+\Sigma_{\rm b}\exp(-r/r_{\rm b})$ to $\Ss$ using a maximum likelihood estimation. Adopting alternative bulge profiles, e.g. with variable S\'ersic index, makes little difference to the quality of these fits and has negligible impact on $\js$. In the fitting procedure, the model $\Sigma_{\rm model,s}$ is blurred by the PSF, which depends on $\x$ due to the combination of adaptive optics and natural seeing.
\item Fit a two-parameter velocity profile\citex{Boissier2003} $v_{\rm model}(r)=v_{\rm flat}(1-\exp[-r/r_{\rm flat}])$ to $v(r)$ using a maximum likelihood estimation.
\end{enumerate}

Using only the first two fitting steps, we compute the directly observed stellar specific angular momentum as
\be\label{eq_jobs}
	\jobs=\Big[\sum_i\Sigma_{s,i}\Big]^{-1}\sum_i\Sigma_{s,i}\,r_i\,v_i,
\ee
where $i$ goes over all pixels with velocity and continuum data, $r_i$ is the radius in the disk plane and $v_i$ is the circular velocity. So far, we only assumed the stellar mass density to be proportional to $\Ss$; the proportionality factor turns out to be irrelevant for $\jobs$ as it cancels out in \eq{jobs}. For the same reason, there is no need to correct $\Ss$ by an inclination factor $\cos^{-1}\incl$. The values $\jobs$ are not converged in the sense that they would increase if data at larger radii were available. To obtain the best estimates of the full $\js$, called $\jextra$, we extend the sums in \eq{jobs} to the domain $\mathbb{R}^2$ (to infinite $r$). Where measurements for $\Ss$ or $v$ are unavailable, their values are approximated by $\Sigma_{\rm model,s}$ and $v_{\rm model}$, respectively.

The best-fit parameters of our targets are listed in Table~1. For C22-2, $r_{\rm flat}$ and $v_{\rm flat}$ are individually almost unconstrained, however, their product is constrained at the 15\% level, such that the rotation curve is well constrained at the radii that most strongly contribute to the angular momentum. We estimate the uncertainties of $\jextra$ by propagating the covariance matrix of the fits, taken as the inverse Hessian of the log-likelihood function. For the two galaxies with both GMOS and OSIRIS data (G04-1, G20-2), we verified that the extrapolations based solely on OSIRIS data ($\jextra=958\pm18\%,\ 397\pm13\%$) are consistent with the more accurate extrapolations from combined OSIRIS+GMOS data ($\jextra=834\pm14\%,\ 394\pm8\%$).

\subsection{C. Tests of the fitting algorithm}

We tested the fitting procedure described in Appendix~B against $10^3$ mock datasets of model-galaxies, as illustrated in Figure~3. The quality of these mock data is similar to the quality of our targets with OSIRIS+GMOS data. We found that our model recovers the true total specific angular momentum $\jextra$ at an accuracy of $\sim\!5\%$ with negligible systematics. Note that the estimated statistical uncertainties of $\jextra$ in Table~1 are generally larger than $5\%$, because the galaxies are not perfectly axially symmetric disks; hence the fits and extrapolations are somewhat less certain.

We also checked the algorithm on the galaxy NGC~3198, a regular $\Mknee$-disk for which kinematic fits\citex{Leroy2008} and $\js$-measurements\citex{Obreschkow2014a} were performed independently at high precision using kinematic data reaching out to $\sim\!10$ exponential radii. For the present test, the data were heavily truncated and smoothed (see Figure~4) to a physical size and resolution similar to the IFS data of the four targets. Applying our fitting algorithm to this degraded version of NGC~3198 produces results in close agreement with those drawn from the reference work\citex{Leroy2008,Obreschkow2014a} (bottom of Table~1).


\begin{figure*}[h]
	\includegraphics[width=0.34\textwidth,trim={57mm 0mm 44mm 20mm},clip]{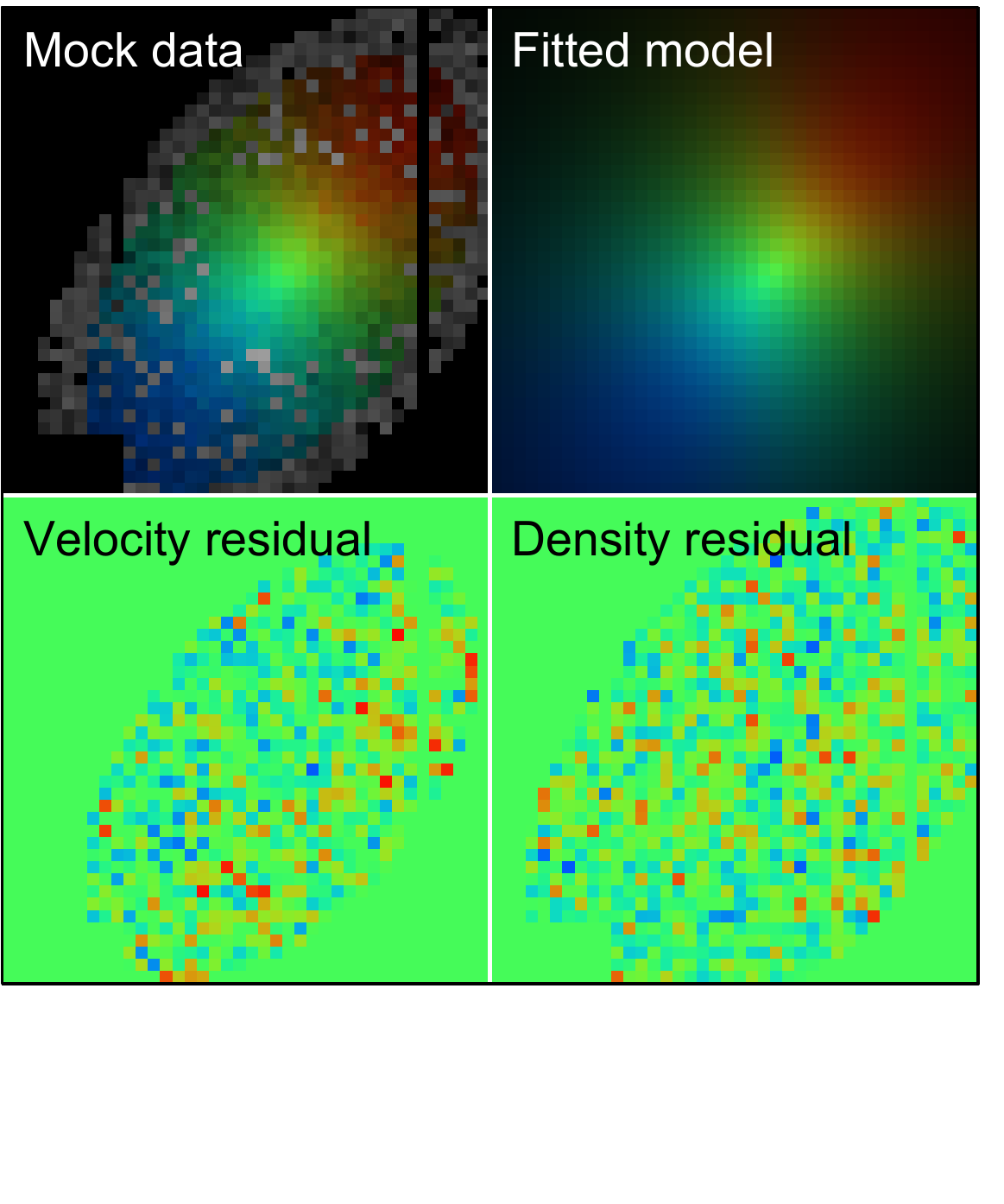}\hspace{5mm}
	\includegraphics[width=0.631\textwidth]{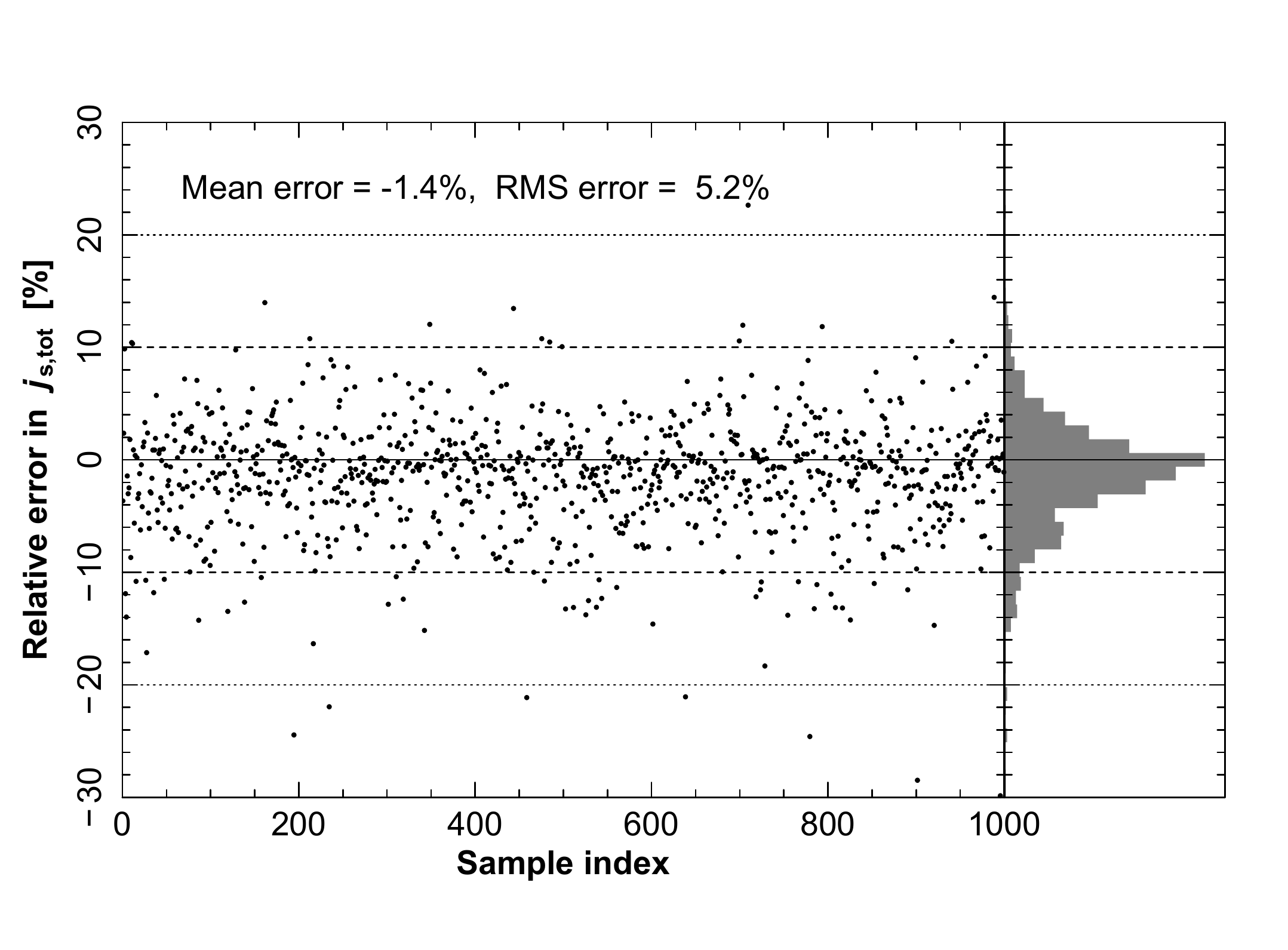}\vspace{-4mm}\\
	{\textbf{Figure 3.} LEFT: example of a mock dataset and the corresponding fits with residuals. The mock data represent an inclined model-galaxy with exponential surface density profile and differential rotation. The data has been degraded in three ways: (1) it is blurred to a physical scale corresponding to the Keck-OSIRIS PSF, (2) signal in some region has been removed (grey and black regions) to mimic sensitivity limitations and sensor errors, and (3) Gaussian noise of realistic amplitude has been added. The colorscales are as in Figure~1. RIGHT: Relative difference between the recovered value of the total specific angular momentum $\jextra$ and the input value in $10^3$ random mock dataset, similar to the one shown in the left panel.\\}
\end{figure*}

\begin{figure*}[h]
	\includegraphics[width=\textwidth]{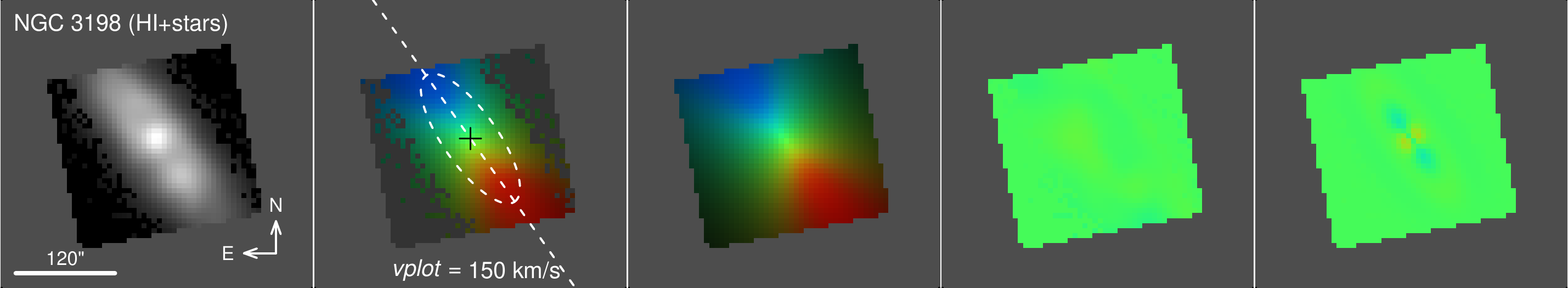}
	\includegraphics[width=\textwidth]{fig_legend.pdf}\vspace{2mm}
	{\textbf{Figure 4.} Same as Figure~1, but for the heavily degraded local control galaxy NGC 3198 (see Appendix~C).}
\end{figure*}




\end{document}